\documentclass[11pt]{article}

\usepackage{epsfig}
\usepackage{array}
\newcommand{\BABARPubYear}{05}

\newcommand{\BABARConfNumber}{003}
\newcommand{\SLACPubNumber}{11314}
\newcommand{\LANLNumber}{0506064}

\input{babarsym}

\newcommand{\DorDstar}{D^{(*)}}
\newcommand{\DorDstarp}{D^{(*)+}}
\newcommand{\DorDstarpm}{D^{(*)\pm}}

\newcommand{\pstar}{p^*}
\newcommand{\vecpstar}{\mathbf{p}^*}
\newcommand{\Estar}{E^*}
\newcommand{\pstarl}{\pstar_{\ell}}

\newcommand{\BY}{\theta_{BY}}
\newcommand{\BX}{\theta_{B\pi\ell}}
\newcommand{\cosBY}{\cos\BY}
\newcommand{\cosBX}{\cos\BX}

\newcommand{\Bperp}{\phi_{B}}
\newcommand{\RR}{\cos^2\Bperp}
\newcommand{\DB}{\Delta\BR}

\newcommand{\eff}{\varepsilon}
\newcommand{\effMC}{\eff^{\mathrm{MC}}}
\newcommand{\effdata}{\eff^{\mathrm{data}}}

\newcommand{\NBB}{N_{\BB}}
\newcommand{\rhoz}{\rho^0}
\newcommand{\rhom}{\rho^-}

\newcommand{\qqmin}{q^2_{\min}}
\newcommand{\qqmax}{q^2_{\max}}

\newcommand{\staterr}[1]{\pm{#1}_{\mathrm{stat.}}}
\newcommand{\systerr}[1]{\pm{#1}_{\mathrm{syst.}}}
\newcommand{\systerrs}[2]{{}^{+#1}_{-#2}{}_\mathrm{syst.}}

\newcommand{\FFerrs}[2]{{}^{+#1}_{-#2}{}_\mathrm{FF}}
\newcommand{\DBR}{\Delta\BR}
\newcommand{\BtoXulnu}{B\to X_u\lnu}
\newcommand{\lnu}{\ell\nu}

\newcommand{\Gthy}{\Delta\zeta}

\newcommand{\etal}{\textit{et al.}}
\newcommand{\PRL}[3]{Phys.\ Rev.\ Lett.\ \textbf{#1}, #2 (#3)}
\newcommand{\PRD}[3]{Phys.\ Rev.\ D \textbf{#1}, #2 (#3)}
\newcommand{\PLB}[3]{Phys.\ Lett.\ B \textbf{#1}, #2 (#3)}
\newcommand{\NIMA}[3]{Nucl.\ Instrum.\ Meth. A \textbf{#1}, #2 (#3)}
\newcommand{\JHEP}[3]{JHEP #1, #2 (#3)}

\newcolumntype{L}{>{$}l<{$}}
\newcolumntype{R}{>{$}r<{$}}
\newcolumntype{C}{>{$}c<{$}}

\long\def\inst#1{\par\nobreak\kern 4pt\nobreak
    {\it #1}\par\vskip 10pt plus 3pt minus 3pt}

\begin{document}
{\pagestyle{empty}

\begin{flushright}
\babar-CONF-\BABARPubYear/\BABARConfNumber\\
SLAC-PUB-\SLACPubNumber\\
hep-ex/\LANLNumber\\
June 2005\\
\end{flushright}

\par\vskip 5cm

\begin{center}\Large \bf
  Branching Fraction for $\Bz\to\pim\ellp\nu$
  and Determination of \Vub\\
  in $\Upsilon(4S)\to\Bz\Bzb$ Events Tagged by
  $\Bzb\to D^{(*)+}\ellm\nub$
\end{center}
\bigskip

\begin{center}
\large The \babar\ Collaboration\\
\mbox{ }\\
\today
\end{center}
\bigskip \bigskip

\begin{center}
\large \bf Abstract
\end{center}
We report preliminary results from a study of the charmless
exclusive semileptonic decay
$\Bz\to\pim\ellp\nu$ based on the data collected at the $\Upsilon(4S)$
resonance using the \babar\ detector at SLAC\@.
The analysis uses events in which the signal $B$ meson recoils
against a $B$ meson that has been reconstructed in a semileptonic
decay $\Bzb\to D^{(*)+}\ellm\nub$.
We extract the total branching fraction
$\BR(\Bz\to\pim\ellp\nu)=(1.03\staterr{0.25}\systerr{0.13})\times10^{-4}$
and the partial branching fractions in three bins of $q^2$,
the invariant mass squared of the lepton-neutrino system.
From the partial branching fractions and theoretical predictions for
the form factors,
we determine the magnitude of the CKM matrix element $|V_{ub}|$.
We find
$\Vub=(3.3\staterr{0.4}\systerr{0.2}\FFerrs{0.8}{0.4})\times10^{-3}$,
where the last error is due to normalization of the form factor.

\vfill
\begin{center}
Contributed to the 
XXII$^{\rm nd}$ International Symposium on Lepton and Photon Interactions at High~Energies, 6/30--7/5/2005, Uppsala, Sweden
\end{center}

\vspace{1.0cm}
\begin{center}
{\em Stanford Linear Accelerator Center, Stanford University, 
Stanford, CA 94309} \\ \vspace{0.1cm}\hrule\vspace{0.1cm}
Work supported in part by Department of Energy contract DE-AC03-76SF00515.
\end{center}

\newpage
} 

\begin{center}
\small

The \babar\ Collaboration,
\bigskip

B.~Aubert,
R.~Barate,
D.~Boutigny,
F.~Couderc,
Y.~Karyotakis,
J.~P.~Lees,
V.~Poireau,
V.~Tisserand,
A.~Zghiche
\inst{Laboratoire de Physique des Particules, F-74941 Annecy-le-Vieux, France }
E.~Grauges
\inst{IFAE, Universitat Autonoma de Barcelona, E-08193 Bellaterra, Barcelona, Spain }
% ...

\end{center}\newpage

\section{INTRODUCTION}

The success of the $B$ Factories has significantly improved
our knowledge of the \CP\ violation in the quark sector.
In particular, the angle $\beta$ of the Unitarity Triangle
has been measured to a $5\%$ accuracy from time-dependent
\CP\ asymmetries in $b\to\ccbar s$ decays~\cite{sin2beta}.
On the other hand,
experimental determination of the other two angles
and of the lengths of the two sides
(with the third side normalized to unit length)
have yet to achieve comparable precision.
One of the two sides, the one opposite to the angle $\beta$,
is of particular interest.
The uncertainty of this side is dominated by the smallest
element $\Vub$, which is known to about 11\% precision.
Improved determination of $\Vub$ therefore translates directly
to a more stringent test of the Standard Model.

Charmless semileptonic decays of the $B$ mesons provide the
best probe for $\Vub$.
Measurements can be done either exclusively or inclusively,
i.e., with or without specifying the hadronic final state.
Since both approaches suffer from significant theoretical
uncertainties, it is important to pursue both types of
measurements and test their consistency.

The exclusive $B\to X_u\ell\nu$ decay rates
are related to $\Vub$ through form factors.
In the simplest case of $B\to\pi\ell\nu$, the differential
decay rate (assuming massless leptons) is given by
\begin{equation}
  \frac{d\Gamma(\Bz\to\pim\ellp\nu)}{dq^2} =
  2\frac{d\Gamma(\Bp\to\piz\ellp\nu)}{dq^2} =
  \frac{G_F^2\Vub^2}{24\pi^3}|f_+(q^2)|^2p_{\pi}^3,
\end{equation}
where $q^2$ is the invariant-mass squared of the
lepton-neutrino system, and $p_\pi$ is the momentum
of the pion in the rest frame of the $B$ meson.
The form factor (FF) $f_+(q^2)$ is calculated with a variety
of theoretical models.
In this paper, we consider recent calculations
by Ball and Zwicky~\cite{Ball05} based on
light-cone sum rules (LCSR) and
by the HPQCD~\cite{HPQCD04} and FNAL~\cite{FNAL04}
Collaborations based on unquenched lattice QCD (LQCD)\@.
The LCSR and LQCD calculations provide the form factor
with reliable uncertainties only in limited ranges of $q^2$.
It is therefore necessary to extrapolate the calculated
form factor using empirical functions, or to measure
the partial decay rates
$\Delta\Gamma(B\to\pi\ell\nu)$ with appropriate cuts on $q^2$,
typically chosen as $q^2<16\gev^2$ and $q^2>16\gev^2$
for use with the LCSR and LQCD calculations, respectively.

Measurements of the partial branching fractions
$\DBR(B\to\pi\ell\nu)$ have been
reported by CLEO~\cite{CLEOpilnu}, Belle~\cite{Bellepilnu},
and by \babar~\cite{nureco}.
The CLEO and \babar\ measurements reconstructed $B\to\pi\ell\nu$ events
by inferring the neutrino momentum from the missing momentum;
the Belle measurement used $B$ mesons recoiling against another
$B$ meson reconstructed in semileptonic decays.
\babar\ has also reported a measurement of the total branching fraction
$\BR(B\to\pi\ell\nu)$
using the recoil of fully-reconstructed hadronic $B$ decays~\cite{Breco}.

In this paper, we report preliminary results from a study of the
$\Bz\to\pim\ellp\nu$ decay,
using an event sample tagged by
$\Bzb\to\DorDstarp\ellm\nub$ decays.\footnote{%
  Charge-conjugate modes are implied throughout this paper.}
A similar study of the $\Bp\to\piz\ellp\nu$ decay is reported in a
separate paper~\cite{pi0lnu}.

\section{THE \babar\ DETECTOR AND DATA SAMPLES}

This measurement uses the $\epem$ colliding-beam data collected with the
\babar\ detector~\cite{BABAR} at the \pep2\ storage ring.
Charged particles are measured by a combination of five-layer
silicon microstrip tracker and a 40-layer central drift chamber,
both operating in a 1.5\,T magnetic field.
A detector of internally reflected Cherenkov light provides
charged kaon identification.
A CsI(Tl) electromagnetic calorimeter provides photon detection
and, combined with the tracking detectors, electron identification.
The instrumented flux return of the magnet identifies muons by
their penetration through the iron absorber.

The data sample analyzed contains 232 million $\epem\to\BB$ events,
where \BB\ stands for $\Bp\Bm$ or $\Bz\Bzb$.
It corresponds to an integrated luminosity of 
211\invfb\ on the $\Upsilon(4S)$ resonance.
In addition, a smaller sample (22\invfb)
of off-resonance data recorded at approximately $40\mev$ below the resonance 
is used for background subtraction and validation purposes.

We also use several samples of simulated $\epem\to\BB$ events
for evaluating the signal and background efficiencies.
Charmless semileptonic decays $\BtoXulnu$ are simulated as a
mixture of exclusive channels
($X_u = \pi$, $\eta$, $\eta'$, $\rho$, and $\omega$)
based on the ISGW2 model~\cite{ISGW2} and
non-resonant $B\to X_u\lnu$ decays~\cite{DFN}
with hadronic masses above $2m_{\pi}$.
For the signal channels,
we give weights to the simulated events in order to
reproduce the $q^2$ distribution predicted by
the recent LCSR~\cite{Ball05}
and unquenched LQCD~\cite{HPQCD04,FNAL04} calculations
as well as the ISGW2 model.

\section{ANALYSIS METHOD}\label{sec:analysis}

The analysis method we use for event selection and signal yield
extraction has been developed blind,
i.e., without using the signal sample in the data.
The procedure described in the following sections has been chosen
and optimized using MC simulation
to obtain the largest expected statistical significance
of the partial signal yields in the three $q^2$ bins defined
in Section~\ref{ssec:yield}.

The outline of the analysis is as follows:
We look for combinations of a $\Dp$ or $\Dstarp$ meson
and a lepton
($e^-$ or $\mu^-$) that are kinematically consistent with
$\Bzb\to\DorDstarp\ellm\nub$ decays.
For each such $B$ candidate, we define the recoil side as
the tracks and calorimeter clusters that are not associated with
the candidate.
We search in the recoil side for a signature of a $\Bz\to\pim\ellp\nu$ decay.
We take advantage of the simple kinematics of the $\Bz\to\pim\ellp\nu$
process to define discriminating variables, and extract the
signal yield from their distributions in three bins of $q^2$.
Finally we calculate the total and the partial branching fractions
using the signal efficiencies predicted by a Monte Carlo (MC) simulation.
We correct for the data-MC efficiency differences using
a control sample in which both $B$ mesons decay to tagging modes.

\subsection{Event Selection}

We search for candidate \BB\ events in which
one $B$ meson decayed as $\Bzb\to\DorDstarp\ellm\nub$,
where the $\Dstarp$ meson is reconstructed in the
$\Dstarp\to\Dz\pip$ and $\Dp\piz$ channels.
The $D$ mesons are reconstructed in the
$\Dz\to\Km\pip$, $\Km\pip\pim\pip$, $\Km\pip\piz$, $\KS\pip\pim$,
and $\Dp\to\Km\pip\pip$ channels.
The widths of the signal regions around the nominal
$D$-meson masses are between $\pm15\mev$ and $\pm30\mev$,
which correspond approximately to $\pm3\sigma$ of the mass resolution.
We also define sideband regions, evenly split below
and above the signal regions, 
which are used to subtract the combinatorial background.
The sideband regions are chosen to be 1.5 times wider than
the signal regions.
The difference between $\Dstarp$ and $D$ masses must be within $\pm3\mev$ of
the nominal values.

The $\DorDstarp$ candidates are combined with an identified
electron or muon to form a $\DorDstarp\ellm\nub$ candidate.
The lepton must have a center-of-mass momentum\footnote{%
  Variables denoted with a star ($x^*$) are measured in the
  $\Upsilon(4S)$ rest frame; others are in the laboratory
  frame}
$\pstarl>0.8\gev$.
If the $D$ meson was reconstructed with a charged kaon,
the kaon charge and the lepton charge must have the same sign.

For each $\Bzb\to\DorDstarp\ellm\nub$ candidate,
we remove from the event the tracks and the neutral clusters
that make up the $\DorDstarp\ellm\nub$ tag.
We then search for a $\Bz\to\pim\ellp\nu$ candidate
in the remaining part of the event.
We require an identified lepton with $\pstarl>0.8\gev$, 
accompanied by an oppositely charged track that is not
identified as either a lepton or a kaon.
To allow for the $\Bz$-$\Bzb$ mixing,
we do not require the two leptons in a candidate event to
be oppositely charged.

The signal events we try to identify contain two neutrinos,
one from each $B$ decay.
Four-momentum conservation and the invariant masses of the
$B$ mesons and neutrinos provide just the sufficient number of
constraints (8) to determine the event kinematics.
Referring to the $\DorDstarp\ellm$ system as the ``$Y$'' system,
we first calculate the cosine of $\BY$, the angle between $\vecpstar_B$
and $\vecpstar_Y$, as
\begin{equation}
  \cos\BY =
  \frac{2\Estar_B\Estar_Y-m^2_B-m^2_Y}
       {2\pstar_B\pstar_Y}.
  \label{eq:cosby}
\end{equation}
The energy $\Estar_B$ and momentum $\pstar_B$ of the $B$ meson are
known from the beam energies and the $\Bz$ mass.
Equation~(\ref{eq:cosby}) assumes that a $\Bzb\to\DorDstarp\ellm\nub$
decay has been correctly reconstructed, and the only undetected
particle in the final state is the neutrino.
If that is the case, $\cosBY$ should be between $-1$ and $+1$ within
experimental resolution.
If the tag has been incorrectly reconstructed,
Equation~(\ref{eq:cosby}) does not give a cosine of a physical angle,
and $\cosBY$ is distributed more broadly.

Analogously,
we can calculate the cosine of $\BX$,
the angle between $\vecpstar_B$ and $\vecpstar_{\pi\ell}$, as
\begin{equation}
  \cos\BX =
  \frac{2\Estar_B\Estar_{\pi\ell}-m^2_B-m^2_{\pi\ell}}
       {2\pstar_B\pstar_{\pi\ell}}.
  \label{eq:cosbx}
\end{equation}
This variable, again, should be between $-1$ and $+1$ for the
signal events, and distributed broadly for the background.

\begin{figure}[b]\centering
\psfig{file=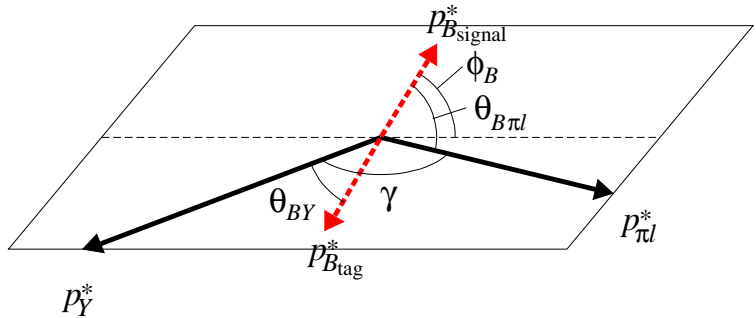,width=0.6\textwidth}
\caption{Angles $\Bperp$, $\BY$, $\BX$, and $\gamma$.
  The $B$ and $\Bbar$ momenta (dashed red arrows) are out of the plane
  defined by the $Y$ and $\pi\ell$ momenta (black arrows).}
\label{fig:phiB}
\end{figure}

The momenta of the two $B$ mesons must be back-to-back
in the center-of-mass frame.
Given $\cosBY$, $\cosBX$, and the directions of
$\vecpstar_Y$ and $\vecpstar_{\pi\ell}$, we can
determine the direction of the $B$ momenta up to a two-fold
ambiguity.
Denoting $\Bperp$ to be the angle between $\vecpstar_B$
and the plane defined by 
$\vecpstar_Y$ and $\vecpstar_{\pi\ell}$, we find
\begin{equation}
  \RR = \frac{\cos^2\BY + \cos^2\BX + 2\cos\BY\cos\BX\cos\gamma}
        {\sin^2\gamma},
  \label{eq:RR}
\end{equation}
where $\gamma$ is the angle between $\vecpstar_Y$
and $\vecpstar_{\pi\ell}$, as shown in Figure~\ref{fig:phiB}.
It should be noted that $\RR$ satisfies
\begin{equation}
  \RR \ge \cos^2\BY\quad\mbox{and}\quad \RR \ge \cos^2\BX
\end{equation}
by construction.
Equation~(\ref{eq:RR}) assumes that $\cosBY$ and $\cosBX$ are correct,
i.e., both the tag-side and the signal-side of the event have been
correctly reconstructed with only one neutrino missing on each side.
In that case, the variable $\RR$ must be between 0 and 1.
If any part of the reconstruction is incorrect, $\RR$ is no longer
a cosine squared of any physical angle, and its distribution spreads
beyond $+1$.

We use $\RR$ as the principal discriminating variable for this analysis.
Since $\RR$ uses the information of the complete event and is strongly
correlated with $\cosBY$ and $\cosBX$, we apply only very loose cuts,
$|\cosBY|<5$ and $|\cosBX|<5$, in the event selection.

If a signal event has been fully reconstructed,
no other particles should be present.
In reality, such an event often contains extra photons,
some of which come from $\piz$s and/or photons
from decays of $\Dstar$ or heavier charmed mesons.
Although we use the $\Dstarp\to\Dp\piz$ decay in the
tags, the efficiency for reconstructing the soft $\piz$
is low.
We do not use the $\Dstarp\to\Dp\gamma$ decay due
to a poor signal-to-background ratio.
We identify the photons that may have come from
$\Dstar\to D\piz$ and $D\gamma$ decays by combining them with 
the $D$ meson candidate; if the combination satisfies
$m_{D\gamma}-m_D<150\mev$ and $\cos\theta_{BY'}<1.1$,
where $Y'$ stands for the $D\gamma\ell$ system, the photon is
considered as a part of the $\DorDstar\lnu$ system and is
removed from the recoil system.
Another source of extra photons is Bremsstrahlung in the
detector material.
If either lepton is an electron,
we identify and remove the Bremsstrahlung
photons based on their proximity to the direction of the electron track.

At this point, we require that the event contains
no charged tracks 
besides the $\Bzb\to\DorDstarp\ellm\nub$ and $\Bz\to\pim\ellp\nu$ candidates.
We further require that there be no residual photons with
laboratory-frame energy $E_{\gamma}>100\mev$.

A few additional cuts are applied to improve the
signal-to-background ratio by rejecting specific types of known
background events.
In order to suppress non-\BB\ background,
we require that the ratio $R_2$ of the second and zeroth Fox-Wolfram
moments~\cite{FW}, computed using all charged tracks and 
neutral clusters in the event, be smaller than 0.5.
The invariant mass of the two leptons, if they are oppositely charged,
must not satisfy $2.95<m_{\epem}<3.15\gev$ or $3.05<m_{\mumu}<3.15\gev$
in order to suppress the background due to $\jpsi\to\ellp\ellm$ decays.
We also calculate the invariant mass of the $\pip\ellm$ system assuming that
the pion was a lepton of the same species as the identified lepton,
and require this mass to be outside 3.06--3.12\gev.

After all the selection cuts,
a small fraction of events contain more than one candidate;
simulated signal events that pass the
selection contain 1.14 candidates on average.
When there are multiple candidates in an event,
we select the candidate with the smallest value of $|\cosBY|$.
The candidates in the $D$-mass sidebands are included in the
selection procedure.

\subsection{Signal Yields}\label{ssec:yield}

We find in the on-resonance data
$966\pm31$ and $725\pm34$ candidate events in the $D$-mass
signal region before and after the sideband subtraction, respectively.
Table~\ref{tab:modes} summarizes the numbers of candidate events
for each decay channel of $\DorDstarp$.
\begin{table}\centering
\caption{\label{tab:modes}
  Numbers of candidate events that enter the signal-yield fit
  from the on-resonance data.
  The $D$-mass sidebands have been subtracted.}
\begin{tabular}{lr@{\,$\pm$\,}r}
\hline\hline
$\DorDstarp$ decay channel & \multicolumn{2}{c}{Events} \\
\hline
$\Dp(\Km\pip\pip)$ & 336.3 & 25.2 \\
$\Dz(\Km\pip)\pip$ & 89.0 & 9.9 \\
$\Dz(\Km\pip\pim\pip)\pip$ & 102.3 & 11.7 \\
$\Dz(\Km\pip\piz)\pip$ & 172.7 & 14.8 \\
$\Dz(\KS\pip\pim)\pip$ & 17.7 & 5.2 \\
$\Dp(\Km\pip\pip)\piz$ & 7.3 & 3.4 \\
\hline
Total & 725.3 & 33.6 \\
\hline\hline
\end{tabular}
\end{table}

\begin{figure}[tb]\centering
\psfig{file=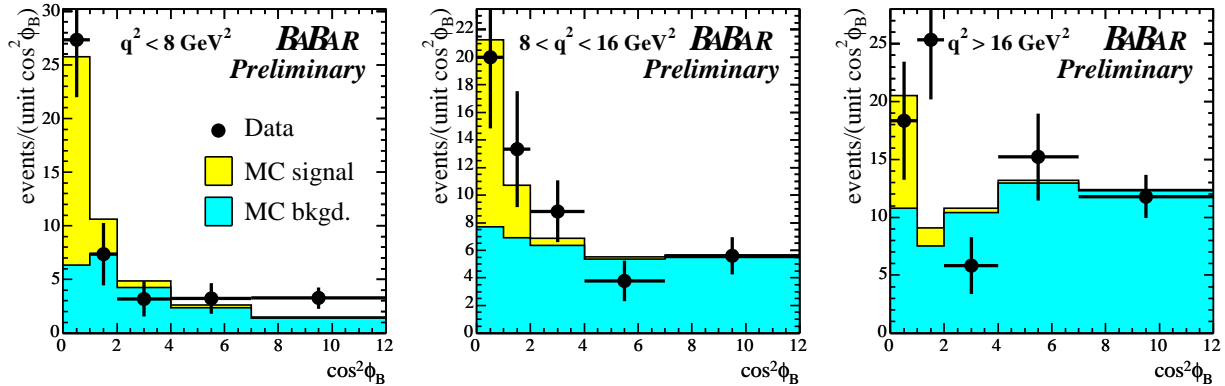,width=\textwidth}
\caption{The $\RR$ fit of the data
  for $q^2<8\gev^2$ (left),
  $8<q^2<16\gev^2$ (middle), and $q^2>16\gev^2$ (right).
  The $m_D$ sidebands have been subtracted.
  The points with error bars are the on-resonance data.
  The histograms are signal (yellow) and $\BB$ background (cyan).}
\label{fig:cfit}
\end{figure}

We extract the signal from the $\RR$ distribution in three
$q^2$ bins, namely, $q^2<8\gev^2$, $8<q^2<16\gev^2$, and $q^2>16\gev^2$,
as shown in Figure~\ref{fig:cfit}.
For each event passing the selection, we calculate
$q^2$ by
\begin{equation}
q^2 = (p_{\ell}+p_{\nu})^2 \approx (\tilde{p}_B-p_{\pi})^2,
  \label{eq:q2}
\end{equation}
where $\tilde{p}_B$ is the approximate $B$ four-momentum defined,
  in the center-of-mass frame, as
\begin{equation}
\tilde{p}^*_B = (\tilde{E}^*_B,\tilde{\mathbf{p}}^*_B)
  \equiv \left(\frac{m_{\Upsilon(4S)}}{2},0,0,0\right).
\end{equation}
In other words, the center-of-mass motion of the $B$ meson is ignored.
Since the $B$ momentum in the center-of-mass frame is small,
the impact of this approximation is small, as will be shown in
Section~\ref{sec:eff}.
It is worth noting that the experimental input to Equation~(\ref{eq:q2})
is the pion momentum and not the lepton momentum.
As a result, electron energy loss due to unrecovered Bremsstrahlung has
no impact on the $q^2$ resolution.

The backgrounds in this measurement are handled in three groups.
First, the combinatoric background for the $D$ mesons is subtracted
using the $D$-mass sidebands.
The remaining backgrounds are mostly \BB\ events,
and are separated from the signal using the $\RR$ distribution.
The possible contribution from the non-\BB\ background is estimated,
based on the $\RR$ distribution of the off-resonance data events
that pass the event selection,
to be smaller than 1.0 event of the total yield,
and is included in the systematic error.

The raw signal yield is extracted in each $q^2$ bin
by a simple binned $\chi^2$ fit of the $\RR$ distribution
of the on-resonance data to the weighted sum of the signal and
\BB\ background distributions from the MC simulation.
The sources of the \BB\ background can be $B\to X_u\lnu$ decays,
$B\to X_c\lnu$ decays, and other (hadronic) $B$ decays.
The $\RR$ distribution for the signal events peaks between
0 and 1, while that of the background is broad with a gradual
fall off toward large values of $\RR$.
The fall-off is faster for smaller $q^2$, as it can be seen in
Figure~\ref{fig:cfit}.
Since the $\RR$ distributions for the \BB\ background from various
sources are quite similar, we fix their relative abundances
in the fit, and later vary them within their systematic uncertainties.
The fit therefore has two free parameters:
the normalization of the signal and
the normalization of the \BB\ background.

To maximize the statistical sensitivity,
the first bin of the $\RR$ histogram should be at least
as narrow as the signal peak.
At the same time, each bin should contain sufficient entries
to prevent the $\chi^2$ fit from becoming biased.
We use variable bin sizes to satisfy the two requirements.
The bin boundaries are chosen to be
$\RR=0$, 1, 2, 4, 7, and 12.

Table~\ref{tab:yield} summarizes the signal yield
obtained from the fit of the on-resonance data,
the $\chi^2$ values (for 3 degrees of freedom) and the corresponding
probabilities.
\begin{table}[tb]\centering
  \caption{Partial signal yields, $\chi^2$ values and probabilities
    in bins of $q^2$.
    The errors are statistical.}
  \label{tab:yield}\smallskip
  \begin{tabular}{ccrl}
  \hline\hline
  $q^2$ bin        & Yield (events) & $\chi^2$ & Prob.\\
  \hline
  $q^2<8\gev^2$    & $26.3\pm8.7$ &  4.6 & 0.20 \\
  $8<q^2<16\gev^2$ & $21.2\pm9.2$ &  1.9 & 0.59 \\
  $q^2>16\gev^2$   & $14.2\pm8.8$ & 11.7 & 0.008 \\
  \hline\hline
  \end{tabular}
\end{table}
The combined $\chi^2$ probability of all three $q^2$ bins is 0.033.
The poor $\chi^2$ value of the last $q^2$ bin is caused by the
large number of events in $1<\RR<2$.
Moving the first bin boundary from 1.0 to 1.5 improves the
$\chi^2$ and increases the yield significantly.
We can find no reason for this behavior other than statistical
fluctuation.
We therefore retain 
the result obtained with the original binning,
which was chosen before the data were unblinded.

\subsection{Signal Efficiencies}\label{sec:eff}

In order to derive the partial branching fractions from the
signal yields, we need the signal efficiency in each $q^2$ bin.
Since the measured value of $q^2$ from Equation~(\ref{eq:q2})
in principle differs from the true value, we define the efficiency
$\eff_{ij}$ as the probability, averaged over electrons and muons,
of a $\Bz\to\pim\ellp\nu$ 
event whose true $q^2$ value belongs to the $j$-th bin
to be found in the $i$-th measured-$q^2$ bin.
With this definition,
the signal yield obtained in the $i$-th reconstructed-$q^2$ bin is
expressed as
\begin{equation}
  N_i = 2\sum_j\eff_{ij}\DB_{j}N_B,
  \label{eq:effmat}
\end{equation}
where the factor of two comes from using both electrons and muons,
$\DB_j$ is the partial branching fraction in the $j$-th $q^2$ bin,
and $N_B$ is the number of $\Bz$ mesons in the
data sample.
Using the $\Upsilon(4S)\to\Bz\Bzb$ branching fraction
$f_{00}=0.488\pm0.013$~\cite{f00}, the number $N_B$
equals $2f_{00}\NBB$,
where $\NBB$ is the number of \BB\ events in the data sample,
and the factor of 2 comes from having two $B$ mesons in each event.

We use the Monte Carlo simulation to estimate $\eff_{ij}$
and correct for the known data-MC differences as
\begin{equation}
N_i = 2\frac{\effdata}{\effMC}\sum_j\effMC_{ij}\DB_{j}N_B.
  \label{eq:effmat2}
\end{equation}
Here
we assumed that a single data-MC efficiency ratio $\effdata/\effMC$
can be applied to all $q^2$ bins.
This is reasonable so long as the ratio is close to unity,
which is in fact the case as it will be shown below.

The efficiency determined from the MC simulation is
\begin{equation}
\effMC_{ij} = \left(\begin{array}{ccc}
    1.142\pm0.065 & 0.050\pm0.011 & 0.004\pm0.004 \\
    0.074\pm0.015 & 1.232\pm0.071 & 0.035\pm0.015 \\
    0.007\pm0.008 & 0.063\pm0.016 & 1.350\pm0.097
    \end{array}\right)\times10^{-3}.
  \label{eq:effmat3}
\end{equation}
The errors are due to Monte Carlo statistics.
The selection efficiency averaged over the three $q^2$
bins is $1.32\times10^{-3}$.
In addition to the experimental resolution, final-state
radiation (FSR) changes the $q^2$ distribution.
All MC samples used in this analysis are generated with
PHOTOS~\cite{PHOTOS} to simulate FSR\@.
The bin-to-bin migration due to FSR is predicted by
the MC simulation to be less than 1.2\%,
and the full size of this effect is included in the
systematic error.

We evaluate the data-MC difference of the $\Bzb\to\DorDstarp\ellm\nub$
selection efficiencies using the double-tag events,
in which both $B$ mesons decay to $\DorDstarpm\ell\nu$.
Properties of the $\DorDstar\lnu$ tags
such as the composition of the $\DorDstar$ decay channels
are similar for the tagged-signal and double-tag events.
The number of double-tag events is proportional to the
square of the tagging efficiency after subtracting the
small contribution from background.

The selection criteria for the double-tag events follow
the main analysis as closely as possible.
In each event, we look for two $\DorDstar\lnu$ tags that do not
share any particles.
We remove all particles that are used in the two tags and
require that there be no charged tracks and
no neutral clusters remaining in the event.

\begin{figure}[b!]\centering
\psfig{file=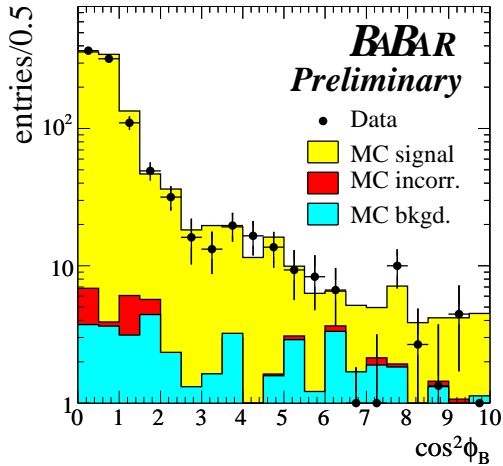,width=0.4\textwidth}
\caption{Distribution of $\RR$ for the double-tag events
  in the data (error bars) and in the MC simulation (histograms).
  The $m_D$ sidebands have been subtracted.
  Colors indicate events with two correct tags (yellow),
  $\BB$ events with two $B\to\DorDstar\ell\nu$ decays that were
  not correctly reconstructed (red),
  and other background (cyan).}
\label{fig:double}
\end{figure}

After subtracting the $D$-mass sidebands,
we find 1073.4 double-tag events in the on-resonance data.
Figure~\ref{fig:double} shows the $\RR$
distribution of the selected events.
The `signal'
in this case consists of the $\Bz\Bzb$ events in which
the two $B$ mesons decay into $\DorDstar\ell\nu$ and are
correctly reconstructed as two tags.
A small fraction of $\Bz\Bzb$ events with two $\DorDstar\ell\nu$
decays are incorrectly tagged, i.e., wrong combinations
of particles are selected as the two tags.
Other sources of background events include
$B\to\tau\to\ell$ cascade decays and
lepton misidentification.

We extract from the study of the double-tag events
the efficiency correction factor
\[
  \frac{\effdata}{\effMC} = 1.000\pm0.047.
\]
The stated error includes both the statistical and systematic uncertainties.
For the latter, we considered
a) the difference in the results when the selection criteria are relaxed
  to allow presence of neutral clusters,
b) the residual background after the $D$-mass sideband subtraction, due
  to possible non-linearities in the backgrounds vs.\ $D$ mass, and
c) the uncertainties in the exclusive $B\to X_c\lnu$ branching
  fractions.
The relaxed criteria used in a) increase the double-tag yield
and the background by approximately 50\% and 100\%, respectively.

\section{SYSTEMATIC UNCERTAINTIES}\label{sec:systematics}

The significant sources of systematic uncertainties 
and their impact on the measured total and partial branching fractions are
summarized in Table~\ref{tab:sys}.
\begin{table}[b!]\centering
\caption{Fractional systematic errors on the total
  and partial branching fractions.}
\label{tab:sys}\smallskip
\begin{tabular}{l|C|CCC}
\hline\hline
& & \multicolumn{3}{c}{$\sigma_{\DB}/\DB$ (\%) in $q^2$ bins} \\
Systematics & \sigma_{\BR}/\BR$ $(\%) & <8\gev^2 & 8$--$16\gev^2 & >16\gev^2 \\
\hline
$B\to\pi\ell\nu$ FF    & +0.2$/$-0.0 & +0.2$/$-0.7 & +0.7$/$-0.3 & +1.6$/$-0.3 \\
$B\to\rho\ell\nu$ FF   & +0.5$/$-0.0 & +0.0$/$-0.2 & +0.5$/$-0.0 & +2.6$/$-0.0 \\
$\BR(B\to\rho\ell\nu)$ & +2.9$/$-3.4 & \pm0.1 & \pm3.7 & +7.8$/$-10.2 \\
$\BR(B\to X_c\ell\nu)$ & \pm2.8 & \pm4.8 & \pm9.9 & \pm13.0 \\
$\BR(B\to X_u\ell\nu)$ & \pm1.7 & \pm0.1 & \pm1.9 & \pm5.6 \\
Shape Function         & +0.8$/$-1.1 & \pm0.1 & \pm0.6 & +2.8$/$-4.1 \\
$\DorDstar\ell\nu$ Tagging & \pm4.7 & \pm4.7 & \pm4.7 & \pm4.7 \\
Tracking               & \pm1.6 & \pm1.6 & \pm1.6 & \pm1.6 \\
Electron ID            & \pm1.1 & \pm1.2 & \pm1.1 & \pm0.9 \\
Muon ID                & \pm1.4 & \pm1.2 & \pm1.4 & \pm1.6 \\
$q^2$ resolution       & $NA$     & \pm2.0 & +0.6$/$-0.8 & \pm2.9 \\
Final-state radiation  & $NA$     & \pm1.2 & \pm1.2 & \pm1.2 \\
$\RR$ for backgrounds  & +8.8$/$-9.1 & +5.6$/$-1.6 & +10.1$/$-14.2 & +14.0$/$-17.9 \\
Non-$\BB$ background   & \pm1.6           & \pm1.6           & \pm1.6             & \pm1.6 \\
Combinatoric $D$       & \pm0.8           & \pm0.8           & \pm0.8             & \pm0.8 \\
$\NBB$                 & \pm1.1 & \pm1.1 & \pm1.1 & \pm1.1 \\
$f_{00}$               & \pm2.7 & \pm2.7 & \pm2.7 & \pm2.7 \\
MC statistics          & \pm3.5           & \pm5.8           & \pm6.1             & \pm7.6 \\
\hline
Total & +12.2$/$-12.5 & +11.5$/$-10.2 & +17.2$/$-19.9 & +24.1$/$-27.4 \\
\hline\hline
\end{tabular}
\end{table}

For the $B\to\pi\lnu$ form factor,
we use the Ball-Zwicky calculation
for our central values, and consider the differences between that and
the two LQCD calculations as the systematic uncertainties.
The form factor affects the branching fractions through the
$q^2$ dependence of the signal efficiency.
As a result, only the shape and not the normalization of the
form factor is relevant at this stage, while the normalization becomes
important in the determination of $\Vub$ as discussed in
Section~\ref{sec:physics}.
The $B\to\rho\lnu$ decays are significant sources of background
at large $q^2$.
We vary the branching fractions as
$\BR(\Bz\to\rhom\ellp\nu)=(2.69^{+0.74}_{-0.77})\times10^{-4}$ and
$\BR(\Bp\to\rhoz\ellp\nu)=(1.45^{+0.40}_{-0.41})\times10^{-4}$,
based on the measurements~\cite{rholnu} and isospin symmetry.
We also compare results using the $B\to\rho\ell\nu$ form factors
calculated by Ball and Zwicky~\cite{Ball05},
Melikhov and Stech~\cite{Melikhov}, and UKQCD~\cite{UKQCD98}.

The branching fractions for the $B\to X_c\lnu$ and $B\to X_u\lnu$ decays
also significantly affect the backgrounds.
We use the latest measurements~\cite{PDG04} for the branching fractions.
Where appropriate, we combine the $\Bz$ and $\Bp$ branching fractions
assuming isospin symmetry.
The shape function parameters used for the simulation of the
non-resonant $B\to X_u\lnu$ decay are varied according to Ref.~\cite{BelleSF}.

The efficiency for the $\DorDstar\ell\nu$ tagging has been discussed
in Section~\ref{sec:eff}.
The systematic uncertainties in the track reconstruction and lepton
identification efficiencies have been derived from studies of independent
control samples.
We vary the amount of migration between the $q^2$ bins due to resolution, 
given by the off-diagonal components in Equation~(\ref{eq:effmat3}),
by $\pm50\%$.
We assign an additional $\pm1.2\%$ error on the partial branching fractions
for the $q^2$-bin migration due to final-state radiation,
which is simulated using PHOTOS~\cite{PHOTOS}.

The largest source of systematic error is the
shape of the $\RR$ distribution for the \BB\ background events.
We studied it using several control samples that are depleted of
the $\pi\lnu$ signal, obtained, for example, by requiring one 
extra charged track or neutral cluster remaining in the event.
The $\RR$ distributions of the control samples agree between the
data and the MC simulation within the available statistics.
We assign the systematic errors based on the statistical uncertainties
of these tests.

We used the off-resonance data sample to set an upper limit on the
residual non-\BB\ background.
MC simulation was used to determine the size of the combinatoric
background that remains after the $D$-mass sideband subtraction,
due to the non-linearities in the backgrounds vs.\ reconstructed
$D$ mass.

The number of $\BB$ events in the on-resonance data sample 
is known to $\pm1.1\%$.
We use $f_{00}=0.488\pm0.013$~\cite{f00} as
the $\Upsilon(4S)\to\Bz\Bzb$ branching fraction.
Limited statistics of the MC samples affects the
measurement primarily through the estimation of the 
signal efficiency $\effMC_{ij}$ given in Equation~(\ref{eq:effmat3}).

In addition to the studies discussed above,
we perform a large number of crosschecks and tests of cut-value dependences.
We investigate all cases in which the variations exceed the expected
statistical fluctuations, and find no indications of systematic problems.

\section{RESULTS}\label{sec:physics}

From the signal yields and the efficiencies evaluated in Section~\ref{sec:analysis},
we extract the following preliminary result for the total branching fraction:
\[
  \BR(\Bz\to\pim\ellp\nu) = (1.03\staterr{0.25}\systerr{0.13})\times10^{-4}.
\]
The partial branching fractions and their
statistical errors are given in Table~\ref{tab:bfresult}.
\begin{table}\centering
  \caption{Preliminary results for the partial and total
    $\Bz\to\pim\ellp\nu$ branching fractions.
    The errors are statistical.}
  \label{tab:bfresult}\smallskip
  \makebox[\textwidth]{%
  \begin{tabular}{l|CCC|C}
  \hline\hline
  & \multicolumn{3}{c|}{$\DB(\Bz\to\pim\ellp\nu)\times10^4$} & \\
  Signal FF   & <8\gev^2 & 8$--$16\gev^2 & >16\gev^2 & \BR\times10^4 \\
  \hline
  Ball-Zwicky~\cite{Ball05} & 0.483\pm0.166 & 0.337\pm0.162 & 0.209\pm0.140 & 1.029\pm0.253 \\
  HPQCD~\cite{HPQCD04}      & 0.480\pm0.165 & 0.339\pm0.163 & 0.211\pm0.142 & 1.031\pm0.254 \\
  FNAL~\cite{FNAL04}        & 0.483\pm0.166 & 0.337\pm0.162 & 0.209\pm0.141 & 1.029\pm0.253 \\
  ISGW2~\cite{ISGW2}        & 0.477\pm0.165 & 0.339\pm0.163 & 0.215\pm0.147 & 1.032\pm0.256 \\
  \hline\hline
  \end{tabular}}
\end{table}
Note that the errors in the partial branching fractions are
negatively correlated because of the small migration across
$q^2$ bins, so adding them in quadrature does not give the
error in the total branching fraction.
We extracted the branching fractions for each of four signal
form-factor calculations
(Ball-Zwicky, HPQCD, FNAL, ISGW2);
the differences are small.

\begin{figure}[tb]\centering
\psfig{file=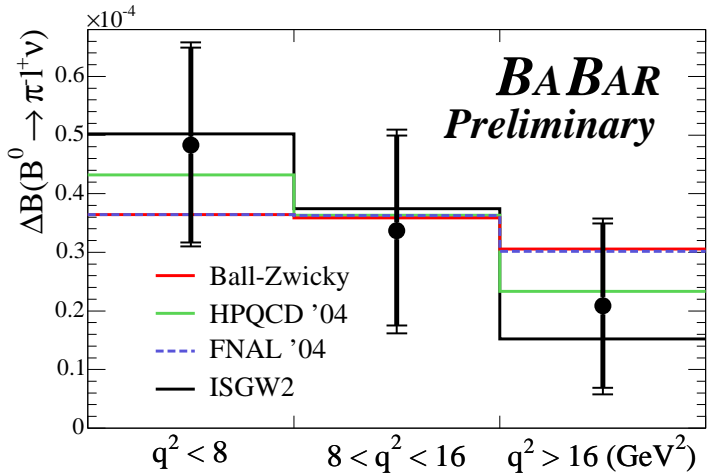,width=0.6\textwidth}
\caption{The partial branching fractions $\DB(\Bz\to\pim\ellp\nu)$.
  The points with error bars are the measurements; the thick parts
    of the error bars indicate the statistical errors.
  The histograms are predictions by  Ball-Zwicky~\cite{Ball05} (red),
  HPQCD~\cite{HPQCD04} (green), FNAL~\cite{FNAL04} (dashed blue),
  and ISGW2~\cite{ISGW2} (black) calculations,
  each scaled to the measured total branching fraction.
  Thus only the $q^2$ dependences are relevant in this comparison.
  The Ball-Zwicky (red) and FNAL (dashed blue) lines overlap with each other.}
\label{fig:cdb}
\end{figure}

Figure~\ref{fig:cdb} compares the measured partial branching fractions
with the $q^2$ dependence predicted by Ball-Zwicky, HPQCD, FNAL,
and ISGW2 calculations.
The calculations are normalized to the measured total branching
fraction.
Although the measured $\DB$ values depend slightly on the form-factor
calculation, the differences are too small ($<2.5\%$)
to be noticeable on this plot.

Table~\ref{tab:babarbf} summarizes the measurements of $\BR(B\to\pi\lnu)$
by the \babar\ collaboration.
\begin{table}[tb]\centering
  \caption{\label{tab:babarbf}
    \babar\ measurements of $\BR(B\to\pi\lnu)$.
    All results are preliminary.
    The last row shows this measurement and the $\Bp\to\piz\ellp\nu$
    measurement in Ref.~\cite{pi0lnu}.}
  \begin{tabular}{lCC}
  \hline\hline
  Technique & \BR(\Bz\to\pim\ellp\nu)\times10^4 & \BR(\Bp\to\piz\ellp\nu)\times10^4 \\
  \hline
  Neutrino reco.~\cite{nureco} & 1.41\pm0.17\pm0.20 & 0.70\pm0.10\pm0.10 \\
  Hadronic tag~\cite{Breco}    & 0.89\pm0.34\pm0.12 & 0.91\pm0.28\pm0.14 \\
  Semileptonic tag             & 1.03\pm0.25\pm0.13 & 1.80\pm0.37\pm0.23 \\
  \hline\hline
  \end{tabular}
\end{table}
Assuming isospin symmetry and the ratio of $B$ lifetimes
$\tau_{\Bp}/\tau_{\Bz}=1.081\pm0.015$~\cite{PDG04},
the measurements agree with each other
with $\chi^2=10.3$ for 5 degrees of freedom,
which corresponds to a one-sided probability of 7\%.  

From the measurement of the partial branching fractions $\DB$
described in this paper,
we extract \Vub\ using
\begin{equation}
\Vub = \sqrt{\frac{\DB}{\Gthy\cdot\tau_{\Bz}}},
\end{equation}
where $\tau_{\Bz} = 1.536\pm0.014\ps$~\cite{PDG04} is the $\Bz$ lifetime,
and $\Gthy$ is defined as
\begin{equation}
\Gthy = \frac{G_F^2}{24\pi^3}\int_{\qqmin}^{\qqmax}|f_+(q^2)|^2p_{\pi}^3dq^2.
\end{equation}
To minimize the theoretical error on \Vub, the range of $q^2$ should 
correspond to the region in which the form-factor calculation is most 
reliable: $q^2<16\gev^2$ for LCSR and $q^2>16\gev^2$ for LQCD.
In order to extract \Vub\ from the total branching fraction $\BR$,
the form factor must be extrapolated to the full range of $q^2$.
This is done in Refs.~\cite{Ball05,HPQCD04,FNAL04} using empirical
functions, and additional uncertainties are assigned for the
extrapolation.

Table~\ref{tab:vub} summarizes the values of $\Vub$ extracted
from the measured partial and total branching fractions.
\begin{table}[tb]\centering
  \caption{\label{tab:vub}
    Preliminary results of \Vub\ extracted from the measured partial
    (first three rows) and total (last three rows) branching fractions 
    and form-factor calculations.}
  \begin{tabular}{lccc}
    \hline\hline
    FF calculation & $q^2$ range & $\Gthy$ ($\ps^{-1}$) & \Vub ($10^{-3}$) \\
    \hline
    Ball-Zwicky~\cite{Ball05}
    & $<16\gev^2$ & $5.44\pm1.43$
    & $3.1\staterr{0.4}\systerr{0.2}\FFerrs{0.5}{0.3}$ \\
    HPQCD~\cite{HPQCD04}
    & $>16\gev^2$ & $1.29\pm0.32$
    & $3.3\staterr{1.1}\systerr{0.5}\FFerrs{0.5}{0.3}$ \\
    FNAL~\cite{FNAL04}
    & $>16\gev^2$ & $1.83\pm0.50$
    & $2.7\staterr{0.9}\systerr{0.4}\FFerrs{0.5}{0.3}$ \\
    \hline
    Ball-Zwicky~\cite{Ball05}
    & full & $7.74\pm2.32$
    & $2.9\staterr{0.4}\systerr{0.2}\FFerrs{0.6}{0.4}$ \\
    HPQCD~\cite{HPQCD04}
    & full & $5.70\pm1.71$
    & $3.4\staterr{0.4}\systerr{0.2}\FFerrs{0.7}{0.4}$ \\
    FNAL~\cite{FNAL04}
    & full & $6.24\pm2.12$
    & $3.3\staterr{0.4}\systerr{0.2}\FFerrs{0.8}{0.4}$ \\
    \hline\hline
  \end{tabular}
\end{table}
The last errors on $\Vub$ come from the uncertainties in $\Gthy$, which
in turn come from the uncertainties of the normalization
of the form-factor calculations.
The precision of the results obtained using the LQCD
calculations~\cite{HPQCD04,FNAL04} in $q^2>16\gev^2$ are
limited by the large statistical error
of the measured partial branching fraction.
Using the total branching fraction reduces experimental
errors on $\Vub$ at the cost of increased theoretical
uncertainties due to the extrapolation of the form factor.

Instead of averaging results based on different theoretical calculations,
we report the value of $\Vub$ obtained from the total branching fraction
based on one of the LQCD calculations~\cite{FNAL04},
\[
\Vub = (3.3\staterr{0.4}\systerr{0.2}\FFerrs{0.8}{0.4})\times10^{-3}.
\]
as a representative result.
This result lies between the results based on the other two calculations,
and carries the most conservative theoretical uncertainty.

\section{SUMMARY}\label{sec:summary}

Using event samples tagged by $\Bzb\to\DorDstarp\ellm\nub$ decays, we obtain
the exclusive branching fraction $\BR(\Bz\to\pim\ellp\nu)$.
The preliminary result for the total branching fraction is
\[
  \BR(\Bz\to\pim\ellp\nu) = (1.03\staterr{0.25}\systerr{0.13})\times10^{-4}.
\]
We also obtain the partial branching fractions in three bins of
$q^2$ of the lepton-neutrino pair.
The preliminary results are
\[
\Delta\BR(\Bz\to\pim\ellp\nu)
= \left\{\begin{array}{ll}
  (0.48\staterr{0.17}\systerrs{0.06}{0.05})\times10^{-4} &q^2<8\gev^2,\\
  (0.34\staterr{0.16}\systerrs{0.06}{0.07})\times10^{-4} &8<q^2<16\gev^2,\\
  (0.21\staterr{0.14}\systerrs{0.05}{0.06})\times10^{-4} &q^2>16\gev^2.
  \end{array}\right.
\]
Using the measured total branching fraction and a form-factor
calculation based on lattice QCD~\cite{FNAL04},
we extract
\[
\Vub = (3.3\staterr{0.4}\systerr{0.2}\FFerrs{0.8}{0.4})\times10^{-3}.
\]
where the last error is due to normalization of the form-factor.
We also use other recent calculations of the form factor~\cite{Ball05,HPQCD04}
and find values of $\Vub$, given in Table~\ref{tab:vub},
that are consistent within the experimental and theoretical uncertainties.

\section{ACKNOWLEDGMENTS}
We are grateful for the 
extraordinary contributions of our \pep2\ colleagues in
achieving the excellent luminosity and machine conditions
that have made this work possible.
The success of this project also relies critically on the 
expertise and dedication of the computing organizations that 
support \babar.
The collaborating institutions wish to thank 
SLAC for its support and the kind hospitality extended to them. 
This work is supported by the
US Department of Energy
and National Science Foundation, the
Natural Sciences and Engineering Research Council (Canada),
Institute of High Energy Physics (China), the
Commissariat \`a l'Energie Atomique and
Institut National de Physique Nucl\'eaire et de Physique des Particules
(France), the
Bundesministerium f\"ur Bildung und Forschung and
Deutsche Forschungsgemeinschaft
(Germany), the
Istituto Nazionale di Fisica Nucleare (Italy),
the Foundation for Fundamental Research on Matter (The Netherlands),
the Research Council of Norway, the
Ministry of Science and Technology of the Russian Federation, and the
Particle Physics and Astronomy Research Council (United Kingdom). 
Individuals have received support from 
CONACyT (Mexico),
the A. P. Sloan Foundation, 
the Research Corporation,
and the Alexander von Humboldt Foundation.

\end{document}